\documentclass[%
 reprint,
 amsmath,amssymb
 aps,
]{revtex4-2}

\usepackage{graphicx}% Include figure files
\usepackage{dcolumn}% Align table columns on decimal point
\usepackage{bm}% bold math
\usepackage{color}
\usepackage{wasysym}
\usepackage{epstopdf}
\usepackage{verbatim}
\usepackage{natbib}
\usepackage{epsfig}
\begin{document}

\preprint{APS/123-QED}

\title{Enhancing fiber atom interferometer by in-fiber laser cooling}% Force line breaks with \\

\author{Yu Wang$^1$, Shijie Chai$^1$, Thomas Billotte$^2$, Zilong Chen$^1$, Mingjie Xin$^1$, Wui Seng Leong$^1$, Foued Amrani$^2$, Benoit Debord$^2$, Fetah Benabid$^2$}
\author{Shau-Yu Lan$^1$}%
 \email{sylan@ntu.edu.sg}
 \affiliation{%
$^{1}$Division of Physics and Applied Physics, School of Physical and Mathematical Sciences, Nanyang Technological University, Singapore 637371, Singapore\\
$^{2}$GPPMM Group, XLIM Institute, CNRS UMR 7252, University of Limoges, Limoges, 87060, France}


%\affiliation{
%\affiliation{
 %Third institution, the second for Charlie Author
%\affiliation{%


\date{\today}% It is always \today, today,

\begin{abstract}
We demonstrate an inertia sensitive atom interferometer optically guided inside a 22-cm-long negative curvature hollow-core photonic crystal fiber with an interferometer time of 20 ms. The result prolongs the previous fiber guided atom interferometer time by three orders of magnitude. The improvement arises from the realization of in-fiber $\Lambda$-enhanced gray molasses and delta-kick cooling to cool atoms from 32 $\mu$K to below 1 $\mu$K in 4 ms. The in-fiber cooling overcomes the inevitable heating during the atom loading process and allows a shallow guiding optical potential to minimize decoherence. Our results permit bringing atoms close to source fields for sensing and could lead to compact inertial quantum sensors with a sub-millimeter resolution.
\end{abstract}

\pacs{Valid PACS appear here}% PACS, the Physics and Astronomy
\maketitle

Atom interferometric sensors use optical pulses along atoms' trajectories to split, deflect and recombine two interferometer arms. While large scale free space interferometers have shown unprecedented sensitivity in measuring gravity, inertial sensing, and test of fundamental physics \cite{Cro,Bon,Tin}, the apparatus that is used to house atoms typically has a cross section of tens of centimeters, set by diffraction of the laser beams that are used to interact with atoms. Shrinking the apparatus size could lead to a compact device and allow the atoms to gain proximity to a source of fields of interest to enhance the signal to be detected.

In free space, reducing the laser beam waist comes at the cost of shortening the distance that atoms can effectively interact with the interferometer beams, thus decreasing the interferometer's sensitivity. Alternatively, hollow-core fibers offer a sub-millimeter enclosure that can guide the interferometer beams over diffraction free and configurable paths. However, most free space high sensitivity interferometers require preparation of ultra-cold atoms at sub-$\mu$K temperature in a low noise environment, and strategies to create such conditions for fiber atom interferometers remain to be developed.

\begin{figure}[h]
\includegraphics[scale=0.38]{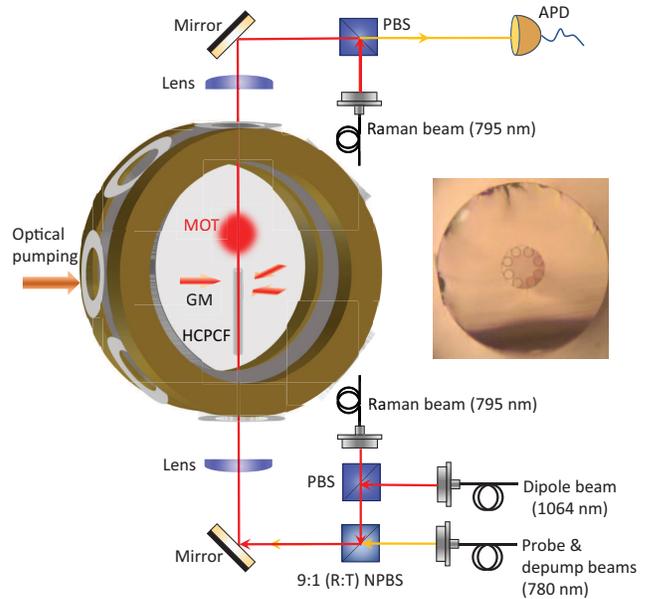}
\caption{\label{fig:Fig1} Experimental setup. HCPCF: hollow-core photonic crystal fiber. PBS: polarization beamsplitter. NPBS: non-polarizing beamsplitter. GM: gray molasses beams. The optical dipole, Raman, probe, and depump beams are all coupled into the fiber. The 2D gray molasses and optical pumping beams are aligned at 5 mm below the upper fiber tip. APD: avalanche photodetector. The inset shows the structure of the fiber used.}
\end{figure}

For example, to guide a large number of the atom into a fiber and avoid their collision with the fiber wall during interferometer sequence, a deep trapping potential is necessary but generates inevitable heating on atoms during loading and guiding \cite{Wan}. The large trapping potential also introduces decoherence on atoms' internal and external states through differential ac Stark shift and inhomogeneous dipole potential along the axial direction, limiting the fiber interferometer time to tens of $\mu$s \cite{Xin}. Using only a few recoil energies of the trapping potential, the interferometer time of optically trapped atoms has reached sub-second in free space \cite{Zha} and 20 s in an optical cavity \cite{Xu}. It is then necessary to cool atoms directly inside the fiber and lower the trapping potential to minimize the decoherence.

Unlike free space laser cooling, the photonic structure of the fiber poses complications on cooling atoms in terms of cooling lasers' geometry and polarization. Although the cooling of atoms along the axial direction of fiber has been demonstrated using Raman sideband cooling \cite{Leo}, the scheme still requires a large trapping potential to confine atoms radially. Here, we implement a two stage cooling of gray molasses and delta-kick cooling, two commonly implemented cooling methods in free-space atom interferometry, to cool the radial temperature of atoms down to 1 $\mu$K. We utilize these cold atoms to demonstrate a Mach-Zehnder interferometer with an interferometer time of 20 ms.

The fiber used in this experiment is a Inhibited-Coupling guiding HCPCF (IC-HCPCF) with a tubular cladding made of eight-tube single-ring \cite{Deb}, as shown in Fig. 1. The fiber core has a diameter of 41 $\mu$m and the 1/$e^{2}$ mode field diameter is 28.7 $\mu$m. We mount a piece of 22-cm-long of the fiber vertically inside an ultra-high vacuum chamber with an angle less than 3$^{\circ}$ from the direction of gravity. Atoms are loaded for 1 s from a two-dimensional magneto-optical trap (2D MOT) into a three dimensional magneto-optical trap (3D MOT) positioned 1.5 cm above the fiber. The temperature of the cold atomic ensemble after sub-Doppler cooling is 8 $\mu$K. After releasing atoms from the 3D MOT, gravity and a dipole force from a 1064 nm laser with power of 700 mW attract the atoms into the fiber.

The optical depth ($OD$) of atoms inside the fiber is determined by measuring the transmission $T_{\textrm{tr}}$ of a probe pulse near resonant on $F=3$ to $F'=4$ transition as $OD=-\textrm{ln}T_{\textrm{tr}}$. Figure 2 shows the resonant $OD$ versus loading time $t$ and the position of atoms relative to the 3D MOT position is plotted as the upper $x$-axis. $OD$ of one corresponds to approximately 6000 atoms in our experimental parameters. The atomic cloud begins to enter the fiber at 25 ms and fully exits the fiber at 210 ms. The major loss mechanism is due to the imperfect coupling of light into the fiber. Unwanted excitation of higher order modes causes modal beating with the fundamental mode, and hence a spatial modulation of the potential along the axial direction \cite{Wan}. The increase of the atom loss rate after 150 ms is mainly due to a bend of the fiber at that position where the dipole force is not strong enough to deflect atoms' trajectory.

We characterize the radial temperature of the atoms inside the fiber using the time-of-flight method \cite{Baj,Xin}. We switch off the dipole beam and let the atomic cloud to expand ballistically for some time $t_{\textrm{r}}$. The probe pulse of 3 nW is then switched on for 50 $\mu$s for the $OD$ measurement as a function of time $t_{\textrm{r}}$, as shown in the inset of Fig. 2. $OD$ with different release time $t_{\textrm{r}}$ can be calculated as

\begin{equation}\label{1}
OD=\frac{OD_{0}}{2r^{2}/W^{2}+1}[1-\exp(\frac{-R^{2}(W^{2}+2r^{2})}{(Wr)^{2}})]
\end{equation}
where $OD_{0}$ is the optical density when $r=0$, $W$ is the 1/$e^{2}$ radius of the guided mode, $r^{2}=r_{0}^{2}+v^{2}t_{\textrm{r}}^{2}$ is the $1/e$ radius of the atomic cloud at time $t_{\textrm{r}}$, $r_{0}$ is the initial radius, $v=2k_{\textrm{B}}T_{\textrm{a}}/m$ is the most probable speed of the atoms, $T_{\textrm{a}}$ is the temperature of atoms, $m$ is the mass, and $R$ is the radius of the inner fiber core. The initial radius of the atomic ensemble in the optical trap under equilibrium is related to the temperature as $r_{0}=W^{2}k_{\textrm{B}}T_{\textrm{a}}/2U$, where $k_{\textrm{B}}$ is the Boltzmann constant and $U$ is the trapping potential. We fit Eq.(1) to the data and obtain the radial temperature of atoms $T_{\textrm{a}}=32(1)$ $\mu$K after loading. The increase of temperature from sub-Doppler cooling temperature is mainly due to the mismatch of the initial atomic cloud distribution and the dipole profile during the loading process.

\begin{figure}
\includegraphics[scale=0.6]{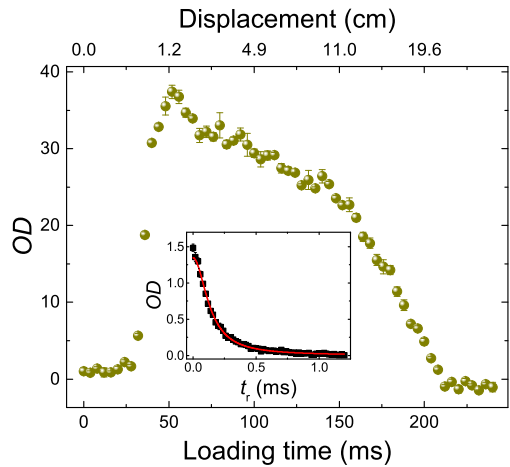}
\caption{\label{fig:2} $OD$ versus loading time after releasing atoms from the MOT. The displacement axis is referenced to the MOT position and calculated using the kinematic equation under free fall and the value of the local gravity 9.78 m/s$^{2}$. The inset shows time-of-flight measurements of atoms along the radial direction at 60 ms after loading, where $t_{\textrm{r}}$ is the time after releasing atoms from the dipole trap.}
\end{figure}

The high temperature of atoms inside the fiber has limited the interferometer time to tens of $\mu$s \cite{Xin}. Here, we apply gray molasses and delta kick cooling to the atoms. The gray molasses serves as a pre-cooling stage for efficient delta-kick cooling. After 60 ms of loading time, the atom cloud enters the fiber at a distance of 2.6 mm from the fiber tip, and we minimize the ambient magnetic field to tens of mG to implement a $\Lambda$-enhanced two-dimensional (2D) gray molasses to cool atoms radially. The 2D gray molasses lasers comprise three laser beams intersecting at 120$^{\circ}$ forming a plane perpendicular to the fiber, as shown in Fig. 1. This laser beam configuration allows to excite the in-fiber atoms thanks to the negligible diffraction off the fiber microstructure. This is enabled by single ring cladding and the absence of photonic bandgap at the beam angle of incidence on the HCPCF. Each beam includes two frequencies $f_{1}$ (pump) and $f_{2}$ (repump) which are +20 MHz detuned from $^{85}$Rb D1 line $F=2$ to $F'=3$ and $F=3$ to $F'=3$ transitions, respectively. When these two frequencies are phase coherent, a dark-state formed by a superposition of $F=2$ and $F=3$ is created to enhance the cooling efficiency \cite{Hsi,Gab}. The power ratio of the two frequencies is 35 to ensure all the atoms are accumulated in the $F=2$ state after 3.2 ms of gray molasses.

\begin{figure*}
\includegraphics[scale=0.82]{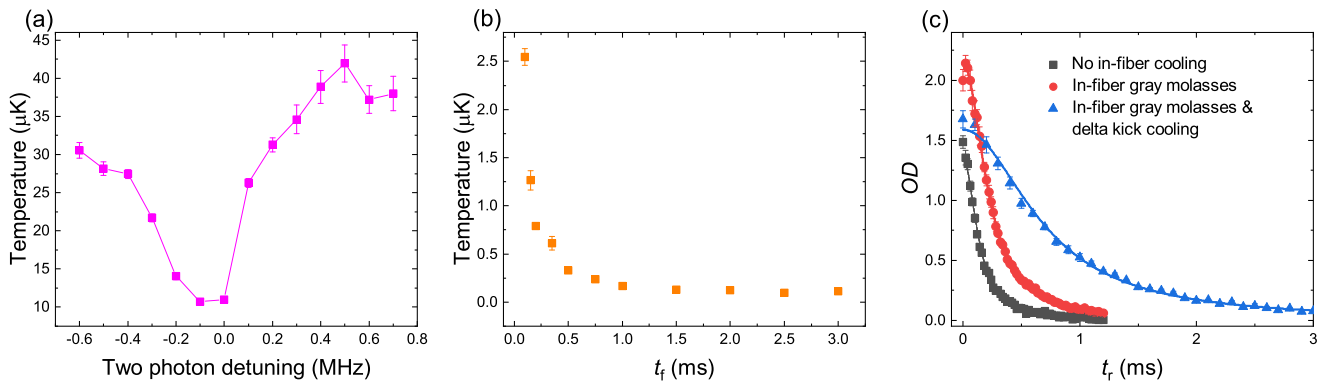}
\caption{\label{fig:Fig1} (a) Measured temperature of atoms in the fiber after gray molasses versus two photon detuning of the pump and repump beams. The lines are drawn to guide the eyes. (b) Measured temperature after delta kick cooling with different free expansion time $t_{\textrm{f}}$. The kicking time $t_{\textrm{k}}$ is determined by the lowest temperature. (c) A comparison of the time-of-flight measurements before and after gray molasses and delta kick cooling. The fitted temperature for these three measurements are 32(1) $\mu$K (black squares), 11.6(2) $\mu$K (red circles), and 1.1(1) $\mu$K (blue triangles), respectively.}
\end{figure*}

Figure 3(a) shows the measured temperature versus two-photon detuning of the molasses beams. The lowest temperature of 10 $\mu$K is achieved at two-photon resonance, a characteristic of the gray molasses involving the dark states from the $\Lambda$ configuration. To study the heating of atoms,  we measure the temperature at a loading time of 160 ms. The temperature increases from 10 $\mu$K after gray molasses at a loading time of 60 ms to 35 $\mu$K. The moderate increase of the temperature excludes atom loss in loading due to the heating of the dipole laser.

To avoid decoherence from any magnetic fields during the interferometer time, atoms are optically pumped into $F$=2 and $m_{F}=0$ state by a $\pi$-polarized optical pumping beam on $F=2$ to $F'=2$ D1 transition incident from the side of the fiber and a linear polarized depump beam on $F=3$ to $F'=3$ D2 transition coupled into the HCPCF core, where $m_{F}$ is the Zeeman state. The quantization axis is defined by a 550 mG magnetic field along the fiber axis. After that, the optical dipole trap is switched off, letting atoms to expand for sometime $t_{\textrm{f}}$. This free expansion creates a correlation between the position and velocity of atoms, where atoms with higher velocity move farther away from the trap center. The trap is then switched on again for a duration $t_{\textrm{k}}$ and atoms are decelerated by the position dependent restoring force from the trap. Ideally, the impulse generated by the restoring force fully stops the momentum of atoms under the condition $t_{\textrm{f}}t_{\textrm{k}}=1/\omega^{2}$, where $\omega=2\pi\times 3.7$ kHz is the radial trap frequency. The temperature $T_{a}$ that can be achieved is determined by the free expansion time as $T_{0}/T_{a}=\omega^{2}t_{\textrm{f}}^{2}$ when $\omega^{2}t_{\textrm{f}}^{2}\gg1$, where $T_{0}$ is the initial temperature \cite{Mar,Kov}.

Figure 3(b) shows the time of flight measurements with different $t_{\textrm{f}}$. The kicking time $t_{\textrm{k}}$ is determined by the best resultant temperature. At $t_{\textrm{f}}=1$ ms, the temperature of 165 nK is optimized at $t_{\textrm{k}}=15$ $\mu$s. The lowest temperature starts to flatten out after $t_{\textrm{f}}=1$ ms. This is mainly due to the anharmonicity of the trap. Moreover, due to the finite size of the guided mode, the long free expansion time for low temperature results in a significant loss of atoms.

Figure 3(c) compares the time-of-flight measurements without cooling, with gray molasses, and with both gray molasses and delta kicking cooling. The decrease of the temperature after gray molasses also increases the density of atoms in the fiber by 30 $\%$ which can be seen from the increase of the initial $OD$ compared to the results without any in-fiber cooling. At $t_{\textrm{f}}=200$ $\mu$s and $t_{\textrm{k}}=24$ $\mu$s, the fitted temperature of the time-of-flight measurements is 1.1(1) $\mu$K, a factor of two larger than the expected temperature of 0.5 $\mu$K. The product $t_{\textrm{f}}t_{\textrm{k}}=4.8\times10^{-9}$ s$^{2}$ is a factor of three larger than 1/$\omega^{2}=1.9\times10^{-9}$ s$^{2}$. These deviations from the ideal scenario also suggest the anharmonicity of the trap when atoms are away from the trap center.

After the delta kick cooling pulse, we reduce the dipole power to 245 $\mu$W (105 nK trapping potential) to weakly trap the atoms and start the interferometer sequence. We implement a Mach-Zehnder interferometer sequence (beamsplitter-mirror-beamsplitter) when atoms are under free-fall. The beamsplitter and mirror optical pulses are formed by a pair of counter-propagating laser beams coupled into the fiber. These two beams drive a two-photon Raman transition on the $F=2, m_{\textrm{F}}=0$ and $F=3, m_{\textrm{F}}=0$ states. The differential ac Stark shift from the dipole beam between these two states is estimated only 60 mHz. To ensure the two frequencies of the Raman beams are phase coherent, a free-running distributed Bragg reflector laser, 20 GHz red-detuned from the $F=2$ to $F'=3$ states D1 line, double-pass through a 1.5 GHz acoustic-optical modulator (AOM) to generate the other frequency of the Raman beam. The zero-order and first-order beams from the 1.5 GHz AOM pass through two separate 80 MHz AOMs to control its intensity, frequency, and phase. The mirror pulse duration of 1.5 $\mu$s corresponds to an effective two-photon Rabi frequency of 2$\pi\times$ 333 kHz, larger than the measured Raman spectrum of 40 kHz of the velocity width of the atoms.

\begin{figure*}
\includegraphics[scale=0.85]{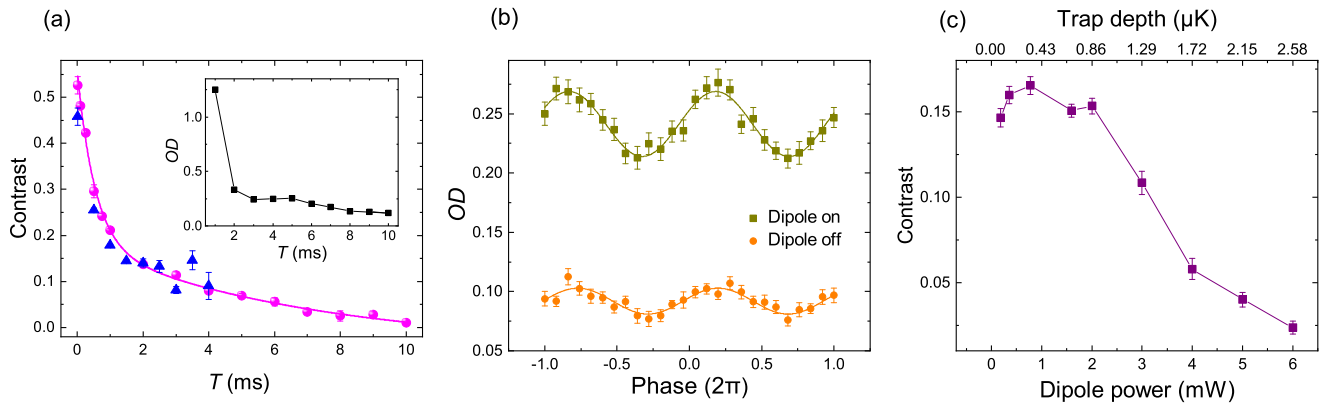}
\caption{\label{fig:Fig5} (a) Contrast of the Mach-Zehnder interferometer versus pulses separation time $T$ with dipole laser on (circles) and off (triangles) during the interferometer sequence. The curve is a fit to the dipole laser on data using sum of two decaying exponential functions. The inset is the mean of $OD$ of the population in $F=3$ state after the interferometer is closed versus $T$ with dipole laser on. The lines are drawn to guide the eyes. (b) Interference fringes with dipole laser on and off at $T=3$ ms. The error bars are the standard error of the mean of 24 experimental runs. (c) Contrast of the interference fringes versus different dipole laser power at $T=2$ ms. The lines are drawn to guide the eyes.}
\end{figure*}

After the first interferometer beamsplitter pulse, the hyperfine spin states of atoms are entangled with their momentum states as $|F=2, m_{\textrm{F}}=0\rangle|p_{0}\rangle+e^{\phi_{1}}|F=3, m_{\textrm{F}}=0\rangle|p_{0}+2\hbar k_{eff}\rangle$, where $p_{0}$ is the initial momentum of atoms, $k_{eff}$ is the effective wavenumber of the two Raman beams in the fiber, and $\phi_{1}$ is the phase of the two Raman beams. The interferometer time is thus affected by the spin and motional coherence. The spin coherence of stationary atoms in the fiber has been demonstrated over 100 ms \cite{Xin2}. In this particular experiment, we measure the spin coherence of atoms under free-fall using spin echo sequence by a pair of co-propagating Raman beams. The measured $1/e$ decay time of the spin echo contrast is 15 ms, limited by the inhomogeneity of the magnetic field defining the quantization axis. The relative momentum of the two arms of the interferometer is mainly influenced by the irregularity of the trapping beams during the interferometer sequence \cite{Zha,Ren}.

The phase at the output of the interferometer is $\phi=-k_{eff}g'T^{2}+\phi_{1}-2\phi_{2}+\phi_{3}$, where $g'$ is the projection of gravity along the fiber axis, $T$ is the separation time of the Raman pulses, and $\phi_{i=1,2,3}$ are the phases of the Raman pulses \cite{Xin}. The population of atoms in the $F=3$ state is $P=(1-\textrm{cos}(\phi))/2$. We vary the phase $\phi_{3}$ of the last Raman pulse and measure the population of atoms in the $F=3$ state to scan the interference fringes.

Figure 4(a) shows the contrast of the interference fringes with different $T$. We observed two different decay trends and fit the data with two exponentially decaying functions. The sharp decay of the contrast in the first 2 ms is due to radial expansion of the atoms in the shallow trap region, which leads to the decrease of the beamsplitter and mirror pulses efficiency. The inset shows the decay of the $OD$ versus $T$ and it matches well with the fast decay trend of the contrast. The second decay trend with fitted decaying constant of 7(3) ms is dominated by the phase fluctuation of the Raman beams from vibrations. To confirm that, we measured the linewidth of the beat note between the two Raman beams to be 150 Hz after passing through the fiber, agreeing well with the contrast decay time we have observed.

The main improvement of our results from \cite{Xin} is the colder temperature of atoms that allows us to use shallower trapping power to reduce the decoherence from the dipole force. To study the influence of the dipole laser only on the coherence, we perform measurements at short $T$ when the dipole laser off during the interferometer sequence. The dipole laser doesn't introduce additional decoherence at $T=3$ ms but retain atoms in the trap, as shown in Fig. 4 (a) and (b). In Fig. 4(c), we study the contrast decay with different dipole power at $T=2$ ms. The contrast starts to decrease at 2 mW of the dipole power due to the irregularity of the dipole beam along the fiber axis.

The initial contrast is limited by the temperature of atoms in the axial direction and the efficiency of the interferometer pulses. By adding gray molasses along the third dimension, a temperature of 5 $\mu$K can be achieved \cite{Ros}. The efficiency of the interferometer pulses can be improved using large bandwidth adiabatic rapid passage \cite{Jaf,Kot}. We expect to improve the interferometer time beyond one hundred ms by installing a vibrational isolation platform to the apparatus and minimizing the higher order modes in coupling the dipole and Raman beams to the fiber.

In summary, we have demonstrated direct laser cooling of atoms inside a hollow-core fiber to below 1 $\mu$K using $\Lambda$-enhanced gray molasses and delta kicking cooling. Both cooling schemes are also applicable to cold atoms trapped and guided by other photonic waveguides, such as nano-fibers and photonic crystal slabs for quantum optics and many body physics experiments \cite{Cha}. With colder atoms in the trap, we extend the coherence time of an inertia sensitive atom interferometer optically guided inside a hollow-core fiber to 20 ms with only 245 $\mu$W of the optical dipole trap and 10 $\mu$W of the Raman beams. The sub-millimeter package of the interferometer could allow short range force and potential measurements with high spatial resolution and can find applications in constraining the deviation of Newton's law of gravity \cite{Bie} and testing the quantum nature of gravity \cite{Car}.

We thank Matt Jaffe, Cris Panda, and Holger M\"{u}ller for discussion. This work is supported by the Singapore National Research Foundation under Grant No. QEP-P4, and the Singapore Ministry of Education under Grant No. MOE2017-T2-2-066.

%\tableofcontents

\nocite{*}

\bibliography{apssamp}% Produces the bibliography via BibTeX.

\end{document}